\documentstyle[12pt,aasms4]{article}

\lefthead{TANIMORI ET AL.}
\righthead{GAMMA RAYS FROM THE CRAB NEBULA}	

\begin{document}

\slugcomment{\sl to be submitted to Astrophysical Journal Letters}

\title{DETECTION OF GAMMA RAYS OF UP TO 50 TeV FROM THE CRAB NEBULA}
\author{T.~Tanimori\altaffilmark{1}, 
K.~Sakurazawa\altaffilmark{1},
S.~A.~Dazeley\altaffilmark{2}, 
P.G.~Edwards\altaffilmark{2,3},
T.~Hara\altaffilmark{4},
Y.~Hayami\altaffilmark{1}, 
S.~Kamei\altaffilmark{1}, 
T.~Kifune\altaffilmark{5}, 
T.~Konishi\altaffilmark{6},
Y.~Matsubara\altaffilmark{7},
T.~Matsuoka\altaffilmark{7},
Y.~Mizumoto\altaffilmark{8}, 
A.~Masaike\altaffilmark{9},   
M.~Mori\altaffilmark{5,10},
H.~Muraishi\altaffilmark{11}, 
Y.~Muraki\altaffilmark{7},
T.~Naito\altaffilmark{12}, 
S.~Oda\altaffilmark{6}, 
S.~Ogio\altaffilmark{1},
T.~Osaki\altaffilmark{1}, 
J.~R.~Patterson\altaffilmark{2},
M.~D.~Roberts\altaffilmark{2,5}, 
G.~P.~Rowell\altaffilmark{2,5}, 
A.~Suzuki\altaffilmark{6},
R.~Suzuki\altaffilmark{1},
T.~Sako\altaffilmark{7}, 
T.~Tamura\altaffilmark{13}, 
G.~J.~Thornton\altaffilmark{2,5},
R.~Susukita\altaffilmark{9,14}, 
S.~Yanagita\altaffilmark{11},
T.~Yoshida\altaffilmark{11}, 
and T.~Yoshikoshi\altaffilmark{1,5} }

\altaffiltext{1}{Department of Physics, Tokyo Institute of Technology, 
 Meguro-ku, Tokyo 152, Japan}
\altaffiltext{2}{Department of Physics and Mathematical Physics, 
 University of Adelaide, South Australia 5005, Australia}
\altaffiltext{3}{Institute of Space and Astronautical Science, Sagamihara, 
 Kanagawa 229, Japan}
\altaffiltext{4}{Faculty of Management Information, Yamanashi Gakuin 
 University, Kofu, Yamanashi 400, Japan}
\altaffiltext{5}{Institute for Cosmic Ray Research, University of Tokyo, 
  Tanashi, Tokyo 188, Japan}
\altaffiltext{6}{Department of Physics, Kobe University, Kobe,
 Hyogo 637, Japan}
\altaffiltext{7}{Solar-Terrestrial Environment Laboratory, Nagoya University,
 Nagoya, Aichi 464, Japan}
\altaffiltext{8}{National Astronomical Observatory, Mitaka,
 Tokyo 181, Japan}
\altaffiltext{9}{Department of Physics, Kyoto University, Kyoto,
 Kyoto 606-01, Japan}
\altaffiltext{10}{Department of Physics, Miyagi University of Education,
 Sendai, Miyagi 980, Japan}
\altaffiltext{11}{Faculty of Science, Ibaraki University, Mito,
 Ibaraki 310, Japan}
\altaffiltext{12}{Department of Earth and Planetary Physics, 
 University of Tokyo, Bunkyo-ku, Tokyo 113, Japan}
\altaffiltext{13}{Faculty of Engineering, Kanagawa University, 
 Yokohama, Kanagawa 221, Japan}
\altaffiltext{14}{Institute of Physical and Chemical Research, Wako, 
Saitama 351-01, Japan}

\authoremail{tanimori@hp.phys.titech.ac.jp}
 
\begin{abstract}
Gamma rays with energies greater than 7 TeV from the Crab pulsar/nebula 
have been observed at large zenith angles,
using the Imaging Atmospheric Technique 
from Woomera, South Australia.
CANGAROO data taken in 1992, 1993 and 1995 indicate
that the energy spectrum extends up to at least 50 TeV,
without a change of the index of the power law spectrum.

The observed differential spectrum is 

\noindent $(2.01\pm 0.36)\times 10^{-13}(E/{7\,\rm TeV})^{-2.53 \pm 0.18}
\,\,{\rm TeV}^{-1}{\rm cm}^{-2}{\rm s}^{-1}$ 
between 7 TeV and 50 TeV.
There is no apparent cut-off.

The spectrum for photon energies above  $\sim$10~TeV 
allows the maximum particle acceleration energy to be inferred, 
and implies that
this unpulsed emission does not originate near the light cylinder of the
pulsar, but in the nebula where the magnetic field is not
strong enough to allow pair creation from the TeV photons.
The hard gamma-ray energy spectrum above 10~TeV also
provides information about the varying role of seed photons for 
the inverse Compton process at these high energies, 
as well as a possible contribution
of $\pi ^{\circ}$-gamma rays from proton collisions.
\end{abstract}


\section{ Introduction }

In the 50~MeV to 10~GeV energy range the Crab emits both pulsed emission,
from the central neutron star, and unpulsed emission,
from the surrounding nebula (Nolan et al.\ 1993).
The Crab nebula has been extensively studied from
radio and optical to X-ray bands,
and these photons are believed to originate as synchrotron emission from
high energy electrons accelerated up to $\sim$ 100 TeV 
in the nebula (De Jager \& Harding \ 1992 hereafter JH92;
Atoyan \& Aharonian \ 1996 hereafter AA96).
The higher energy component, above $\sim$1 GeV, 
however, is thought to be produced by inverse
Compton (IC) scattering between these very high energy electrons 
and ambient photons in the nebula.
In the wide range from GeV to hundreds of TeV,
gamma rays are expected to emanate from the Crab nebula via the IC process.
Spectra have been calculated for various theoretical scenarios
based on the Synchrotron Self Compton (SSC) model, 
using the synchrotron spectrum obtained from
observations of radio to MeV gamma rays as seed photons (JH92;AA96).
The predicted spectra for synchrotron and IC emission are  
sensitive to conditions within the nebula such as 
the magnetic field, the nature of the seed photons, 
and the maximum energy of high energy electrons, 
which allow us to test these models.   
In particular, all models predict that the shape of
the spectrum becomes more sensitive 
to a  change  of the parameters of the nebula
as the gamma-ray energy increases.	

Observations of the IC domain have been made in the
0.1--10 GeV region by {\sl EGRET} 
(Nolan et al.\ 1993; De Jager et al.\ 1996)
and the 0.2--10 TeV region by imaging air \v Cerenkov 
telescopes (IACTs)
(Vacanti et al.\ 1991; Goret et al.\ 1993; Tanimori et al.\ 1994 hereafter T94;
Djannati-Atai et al.\ 1995; Aharonian et al.\ 1997; Cater-Lewis et al.\ 1997).
{\sl EGRET} observations have revealed a 
break in the energy spectrum of the unpulsed component from the Crab nebula
above 0.1 GeV, which is 
believed to signal the change from
synchrotron radiation by several hundreds TeV electrons at lower energies
to IC emission at higher energies.
Unpulsed emission above 200 GeV up to 10 TeV
has been detected by several IACTs 
with high statistics,
although there still remains considerable uncertainty in the absolute
flux and the spectral index.
However, there exists only one observation of the detection for gamma rays 
above $\sim$ 10 TeV with limited statistics: by the Themistocle group 
with a multiple small mirror system (Djannati-Atai et al.\ 1995).

As pointed out above,
the observation of gamma rays in the energy range above 10 TeV 
is the key to understanding of the 
IC process in the nebula.
The large zenith angle technique (Sommers \& Elbert 1987) was used 
by the CANGAROO group in the detection of gamma rays from the Crab in 1992
to obtain a single integral flux point of
$(7.6\pm1.9)\times10^{-13}$ cm$^{-2}$s$^{-1}$ above 7 TeV 
at large zenith angles, $\sim$53\arcdeg (T94), and 
has provided  the current best sensitivity 
for detecting $>$10 TeV gamma rays among the various methods.

\section{Observation}

The observations reported here were made with the 3.8m telescope of 
the CANGAROO collaboration 
(Patterson and Kifune 1992),
which is located at Woomera in South Australia (136\arcdeg47\arcmin E and
31\arcdeg06\arcmin S).
The high resolution camera, set at the prime focus, 
consists of small square-shaped 
photomultiplier tubes of
 10mm $\times$ 10mm size (Hamamatsu R2248).
The number of photomultipliers was  220 in 1993
and was increased to 256, in a 16$\times$16 square pattern, in 1995  
giving a total field of view of about 3\arcdeg.
The details of the camera and telescope are given in Hara et al.\ (1993).
After the observations in 1992,
the Crab was observed again at zenith angles
of 53\arcdeg--56\arcdeg\ in two seasons;
from 1993 December to 1994 January,
and from 1995 December to 1996 January.
In order to monitor the cosmic ray background contained in 'on-source'
data, 'off-source' runs were also done as described 
in T94.
In order to obtain an energy spectrum, 1992 data was 
analyzed again with 1993 and 1995 data. 
The total observation times in those three years 
used in this analysis were $2.22\times10^5$~s 
( $1.36\times10^5$~s for 1993 and 1995 data) for on-source data
and $2.04\times10^5$~s ( $1.20\times10^5$~s for 1993 and 1995 data)
for off source data.

\section{Analysis and Result}

The imaging analysis of the data
is based on parameterization of 
the elongated shape of the \v{C}erenkov
light image as: ``width'',``length'', ``distance'' (location),
``conc'' (shape), and the image orientation angle $\alpha$
(Hillas 1985; Weekes et al.\ 1989).
We analyzed the data of 1992, 1993, and 1995 observations 
by the similar parameters used in T94;
the gamma-ray selection criteria used here
are $0\fdg02 \le$  length $\le 0\fdg33$,  
$0\fdg06 \le$  width $\le 0\fdg13$ and
$0\fdg3 \le$  distance $\le 1\fdg0$.
In addition, the concentration of the yield of \v Cerenkov light
in the image ``conc'' was required, taking account of the
energy dependence.
Here we have determined above cut parameters as loose as possible
to avoid both over-cutting $\gamma$-ray events and the distortion of the 
flatness of the $\alpha$ distribution for background events:
by adjusting a cut parameter not to appear any $\alpha$ peak near 0\arcdeg in
the 'on-source' data removed by the cut,
which eventually reduces the systematic errors. 
The detail is described in T94 .

The resulting event distributions in the combined 1992, 1993 and 1995 
data-set are plotted in Fig.\ref{fig:f1}a as a function of $\alpha$.
The $\alpha$ peak appearing around the origin $(\alpha\le 15\arcdeg)$
in the on-source data
are attributed to $\gamma$-rays  from the Crab, and 
the number of background events in the $\alpha$ peak region
was estimated from the flat region of the $\alpha$ distribution
(30\arcdeg--90\arcdeg) in the on-source data.
Here off-source data was used to verify the non-existence of a peculiar structure
of the $\alpha$ plot around the origin without gamma-ray events.  
The statistical significance of the $\alpha$ peak is calculated 
using the following:
$(N_{\rm on} - \beta \cdot N_{\rm back}) / 
\sqrt{N_{\rm on} + \beta^2 \cdot N_{\rm back}}$,
where  $N_{\rm on}$ and $N_{\rm back}$ are the numbers of events
in the source region  $(0\arcdeg \le \alpha\le 15\arcdeg)$ and
in the background region $(30\arcdeg \le \alpha \le 90\arcdeg)$ 
of the on-source data respectively, 
and $\beta$ is the ratio 
of the $\alpha$ range of the source region (15\arcdeg) 
to the background region (60\arcdeg).
The statistical significance of the   
peak in Fig.\ref{fig:f1}a thus obtained is 7.2$\sigma$.

Effects due to the bright star $\zeta$ Tau, visual magnitude 3.0, 
which is located 1.1\arcdeg\ from
the Crab and within the field of view of the camera, were
investigated by the same procedure 
as used previously (T94). 
We found no effect which would cause false $\alpha$ peaks.
\placefigure{fig:f1}

The collecting  area as a function of energy 
and threshold energy of the telescope for  gamma-ray
showers have been  inferred from a Monte Carlo simulation as described in 
Patterson \& Hillas\ (1990) and T94.
In order to obtain the energy spectrum,
$\alpha$ plots were made by varying the minimum and maximum numbers of 
detected \v Cerenkov  photons,
required in the event analysis.
The energy of gamma-rays is defined as the energy
of the maximum flux of simulated gamma-ray events
having a similar  amount of \v Cerenkov light  to the detected events.
The spectrum was then
estimated as follows.
First, simulated events were generated between 3 TeV and 70 TeV, 
assuming a differential power law index of $-2.6$.
The collecting area, trigger efficiency and 
threshold energy corresponding to each $\alpha$ plot
were independently estimated from the simulation result, and 
a spectrum was calculated.
Using this new spectrum,
the simulation was repeated
and the spectrum recalculated.
We iterated these steps until the spectral index converged.
The resultant differential spectrum, {\it J(E)}, between 7 TeV and 50 TeV
is plotted in Fig.\ref{fig:f2}, 
and can be written as:
\begin{equation}
J(E) = (2.01\pm 0.36)\times 10^{-13}(E/{7\,\rm TeV})^{-2.53 \pm 0.18}
  \,\,{\rm TeV}^{-1}{\rm cm}^{-2}{\rm s}^{-1},
\end{equation}
where the errors are statistical ones only. 

The systematic errors on the flux and the index due mainly 
to the event selection procedure are estimated 
to be $\pm 30$\% and $\pm 0.15$, respectively. 
Another systematic error of 50\% on the flux arises
from the uncertainty of the absolute gamma-ray energy  of 25\%,
which is due mainly to errors in the estimation
of the  number of detected photo-electrons and 
the reflectivity of the telescope mirror.
The total systematic error on the flux is 58\%.

\placefigure{fig:f2}
\placefigure{fig:f3}
The resultant differential spectrum smoothly connects 
with new results of the Whipple group 
from 500~GeV and up to 8~TeV (Cater-Lewis et al.\ 1997) 
as shown in Fig.\ref{fig:f2}.
For comparison with other results above 10 TeV,
the integral flux is presented in Fig.\ref{fig:f3},
which is consistent with
the result of 
the Themistocle Group (Djannati-Atai et al.\ 1995) and several upper
limits of gamma-ray flux obtained by array-type experiments 
(Alexandreas et al.\ 1991; Amenomori et al.\ 1992; Karle et al.\ 1995).

Figs.\ref{fig:f1}b,\ref{fig:f1}c, and \ref{fig:f1}d show
the $\alpha$ plots for the 
events of $\ge$ 20 TeV, $\ge$ 37 TeV, and $\ge$ 47 TeV, respectively.
Note that the $\alpha$ peak above the hadronic background becomes clearer
as the threshold energy increases.
In order to show the hardness of the observed spectrum,
the ratio of the number of gamma rays to that of the background,
are plotted as a function of threshold energy of gamma rays in Fig.\ref{fig:f4},
where gamma-rays events in the $\alpha$ peak ($\alpha \le 15\arcdeg$) 
and background events with  (30\arcdeg$\le \alpha \le$90\arcdeg) were used.
These ratios were calculated using 
the data to which only a simple distance cut were
applied in order to avoid an energy dependence of the rejection efficiencies
of the imaging cuts for hadrons.
The energy dependence of the trigger efficiencies for both gamma rays and
hadrons are taken into account using the same Monte Carlo simulation.
Fig.\ref{fig:f4} shows that this ratio increases 
as the threshold energy rises.
This ratio is expected to be sensitive to the existence of
the spectral cut-off because the hadron spectrum absolutely
extends with the constant index of --2.7 up to more than hundreds TeV. 
Assuming the spectral cut-offs of 30, 50 and 100 TeV,
the energy dependence of this ratio was simulated
including the energy resolution 
as shown in Fig.\ref{fig:f4}, 
where both differential indices
of ours (--2.53) and the Whipple group (--2.45) were examined.
The result obviously favors  the nonexistence of
the spectral cut-off at least less than 50 TeV:
reduced $\chi^2$s of our data are 5.70 (30 TeV cut-off), 
2.74 (50 TeV cut-off) and 0.85 (100 TeV cut-off). 

In addition, the differential spectrum of the background cosmic rays 
was calculated from this data
 to be  
$(3.2\pm 0.1)\times 10^{-8}(E/{10\,\rm TeV})^{-2.73 \pm 0.03}$ 
TeV$^{-1}$str$^{-1}$cm$^{-2}$s$^{-1}$ 
in the energy range of 13~TeV to 110~TeV,
where the error represents only statistical one.
It is noted that the differential cosmic ray spectral index
can be reconstructed with good accuracy,
which also shows the validity of the estimation of the 
relative energy scale for the gamma rays.
The maximum systematic error of our simulation for the
cosmic ray flux is $\sim$50\%
because we assumed all cosmic rays are protons in the simulation.
The resultant flux agrees 
with the recent results of balloon-borne
experiments (Ichimura et al.\ 1993) within $\sim$20\% for the proton flux.
Thus, the spectra of both gamma rays and cosmic rays 
are obtained from the same measurement,
which is the first attempt as far as we know. 
It presents  the independent and absolute calibration 
to estimate the energy spectrum in air \v Cerenkov experiments.

Improvements of the simulation are under way, 
taking account of the other main ingredients of the cosmic ray flux.
\placefigure{fig:f4}

\section{Discussion}

Very high energy gamma rays from the Crab nebula
are believed to originate in the shock acceleration of electrons
by the pulsar wind in the nebula (Kennel \& Coroniti 1984),
but a possible origin within the pulsar magnetosphere  
near the light cylinder (Cheung \& Cheng 1994)
has not been ruled out, observational proof being lacking.
In this region, the strong magnetic 
field decreases the emissivity of
the multi-TeV gamma rays by 
the electron-positron pair creation process
between gamma rays and the magnetic field.
Gamma rays above 10~TeV therefore cannot emanate from
the vicinity of the light cylinder (Cheung \& Cheng 1994).
Our observation of gamma rays up to 50~TeV 
strongly disfavours this idea, 
and provides direct evidence 
that the very high energy relativistic particles are
accelerated by the pulsar wind.

As mentioned previously,
the whole of IC gamma-ray spectrum for the Crab 
have been calculated by 
several authors on the basis of the SSC model. 
Recently it has been pointed out by AA96 that 
Cosmic Microwave Background (CMB) 
and infrared (IR) photons emitted from dust in the nebula
are the main seed photons for TeV gamma-ray production 
by the IC process.
These effects modify the spectrum of TeV gamma rays from the Crab,
causing it to be flatter than that calculated by the simple SSC model.
The spectrum above $\sim$ 100~GeV is
predicted to steepen as the energy increases, 
due to the Klein-Nishina limit in the SSC process.
On the other hand,
the gamma rays produced by IC with low energy seed photons such as
CMB or IR are predicted to have flat spectra 
up to a few TeV (Thomson limit). 
Consequently, the combined spectrum including
the effects of SSC, and IC with CMB and IR,
retains its integral index of $\sim1.50$ up to the $\sim 10$~TeV region.
Observations of TeV gamma rays by the  Whipple, HEGRA,Themistocle, and 
CANGAROO collaborations are evidence of the significant 
contribution of IC with CMB and IR.

Furthermore, as the energy increases,
the observed spectrum departs from the theoretical calculations 
of all models which include only the above contributions such as 
the calculation of AA96.
Intrinsically, the higher energy gamma rays generated by the IC process 
have a naturally steeper spectrum due to synchrotron 
energy loss in magnetic fields. 
Hence all models based on electron processes have
difficulty in explaining the observed hard spectrum 
which extends to beyond 50~TeV.

One simple idea to explain this is to consider the gamma-rays from $\pi^{\circ}$
generated by  protons which may be included in a pulsar wind
and accelerated by its terminated shock similarly to leptons.  
The spectrum of gamma-rays from $\pi^{\circ}$ is expected to retain its 
hard differential index of $\sim-2.1$ up to near the maximum accelerated
energy. 
Above $\sim$10 TeV,
this component gradually may raise its ratio  in the whole gamma-ray spectrum. 
AA96  presented a calculation based on such an assumption
considering  proton acceleration in the nebula,
in which all parameters, such as the total energy used in proton acceleration,
were constrained to satisfy the observational data of the Crab. 
In the case of protons having 
an differential index of $-2.1$, the resulting spectrum 
of TeV gamma rays keeps its differential index of  $-2.5$ beyond 50~TeV which
matches very well with our result.
Although it cannot be said 
that our result is strong evidence of proton acceleration 
in the Crab nebula,
it is supportive of this claim -- further
observations of the Crab are clearly of great importance.

In order to obtain direct evidence of proton acceleration, 
two observations are necessary:
one is the observation of the extension of the hard spectrum up to 100 TeV,
and the other is a measurement of the size of the emission region
of gamma rays $\ge$~10~TeV .
In the Crab nebula, very high energy electrons in the multi-TeV region
emit both X-rays by synchrotron radiation and TeV gamma rays by IC.
The size of the X-ray emission region
is therefore directly related to 
that of the emission region of TeV gamma rays.
The {\sl ROSAT} HRI image at 1--2 keV (Hester et al.\ 1995)
suggests the emission region of TeV gamma rays 
is concentrated within 2\arcmin\ of
the center of the Crab nebula, if TeV gamma rays are generated by IC.
On the other hand, protons can occupy the cavity within the Crab 
nebula of diameter $\sim$ 7\arcmin.
If multi TeV gamma rays from the Crab are mainly due to high energy protons,
the observed source size would increase 
as the energy goes above 10 TeV.

\acknowledgments

T.Tanimori would like to thank F.A. Aharonian for fruitful discussions.
This work is  supported by a Grant-in-Aid in Scientific Research
of the Japan Ministry of Education, Science, Sports  and Culture,
and also by the Australian Research Council and International Science
and Technology Program.
T. Tanimori and T.Kifune are grateful for
support from the Sumitomo Foundation.
The receipt of JSPS Research Fellowships (P.G.E., T.N., M.D.R., K.S.,
G.J.T. and T. Yoshikoshi ) is also acknowledged.

\clearpage

\figcaption[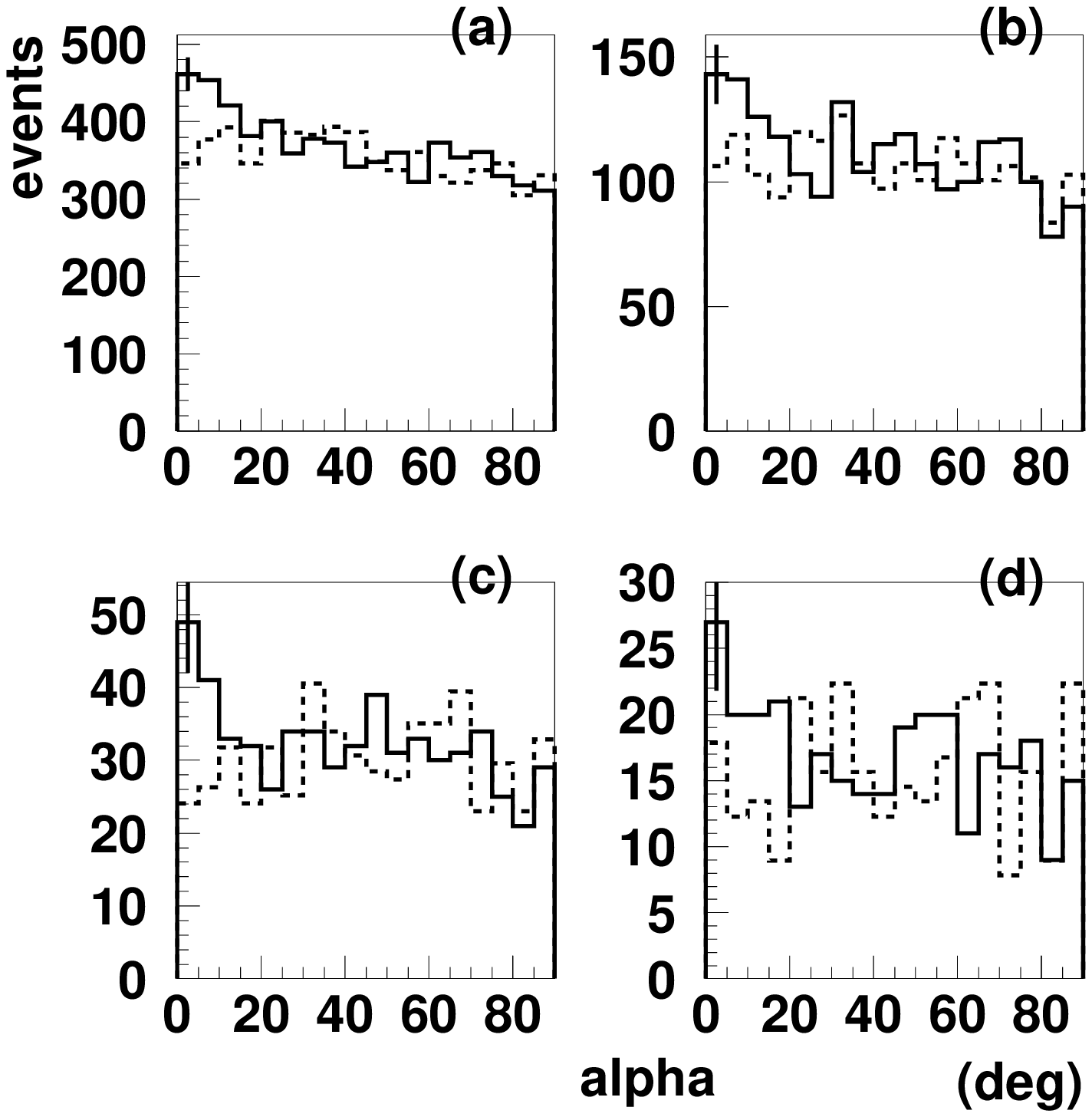]{
(a) The distribution of the image alignment angles relative to 
the source direction, $\alpha$,
for the final combined data (1992, 1993, and 1995):
solid and dashed lines show on- and off-source data, respectively.
Also shown are the $\alpha$ distributions of the same data with 
the threshold energy cuts of (b)$\ge 20$ TeV, (c)$\ge 37$ TeV,
and (d) $\ge 47$ TeV, respectively.
\label{fig:f1}}

\figcaption[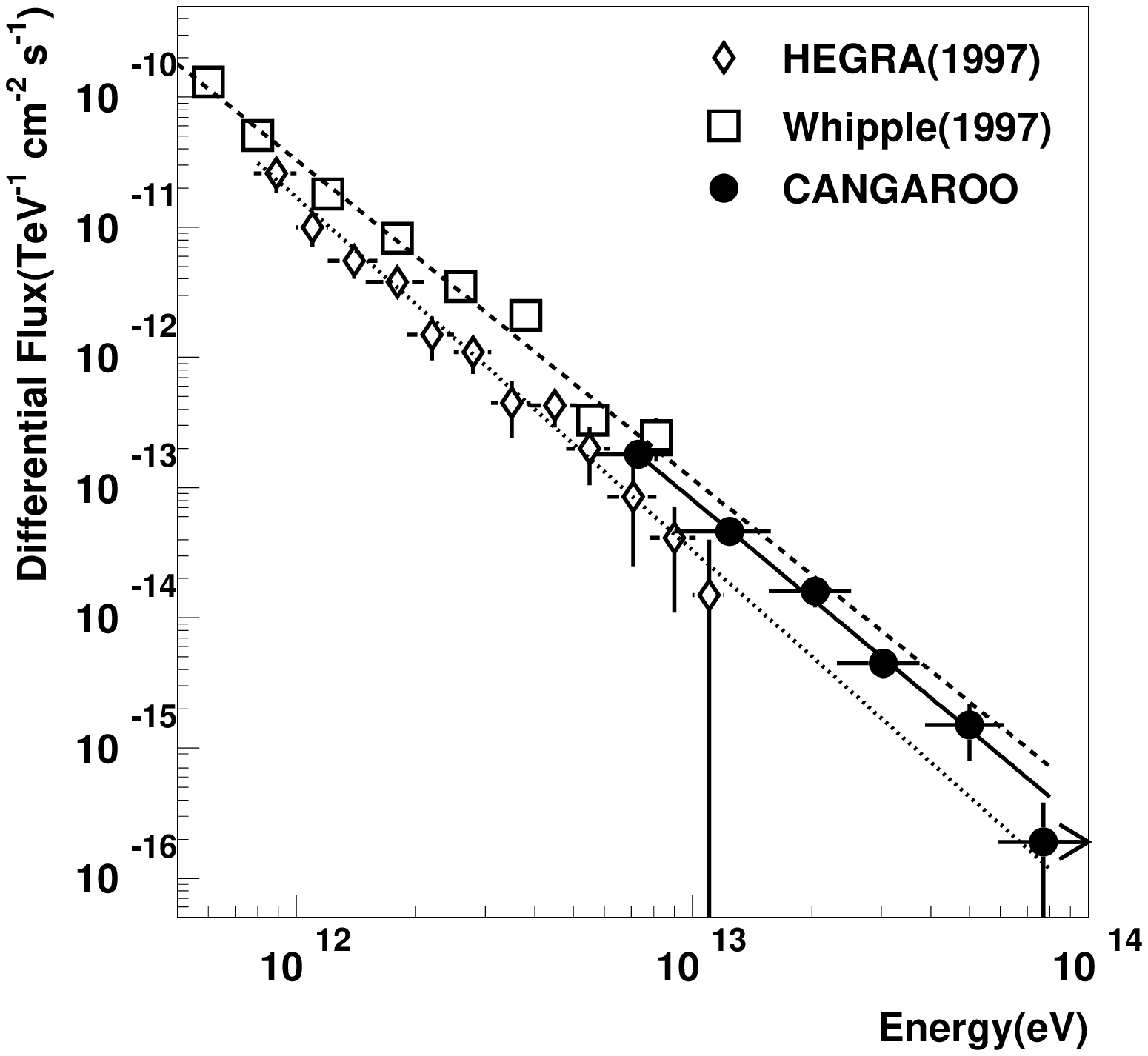]{
The differential  spectra of the present result in comparison with
other experiments:
Recent results from the Whipple (Cater-Lewis et al.\ 1997) and 
the HEGRA groups (Aharonian et al.\ 1997).
The full line is the power law fit given by equation (1).
\label{fig:f2}}

\figcaption[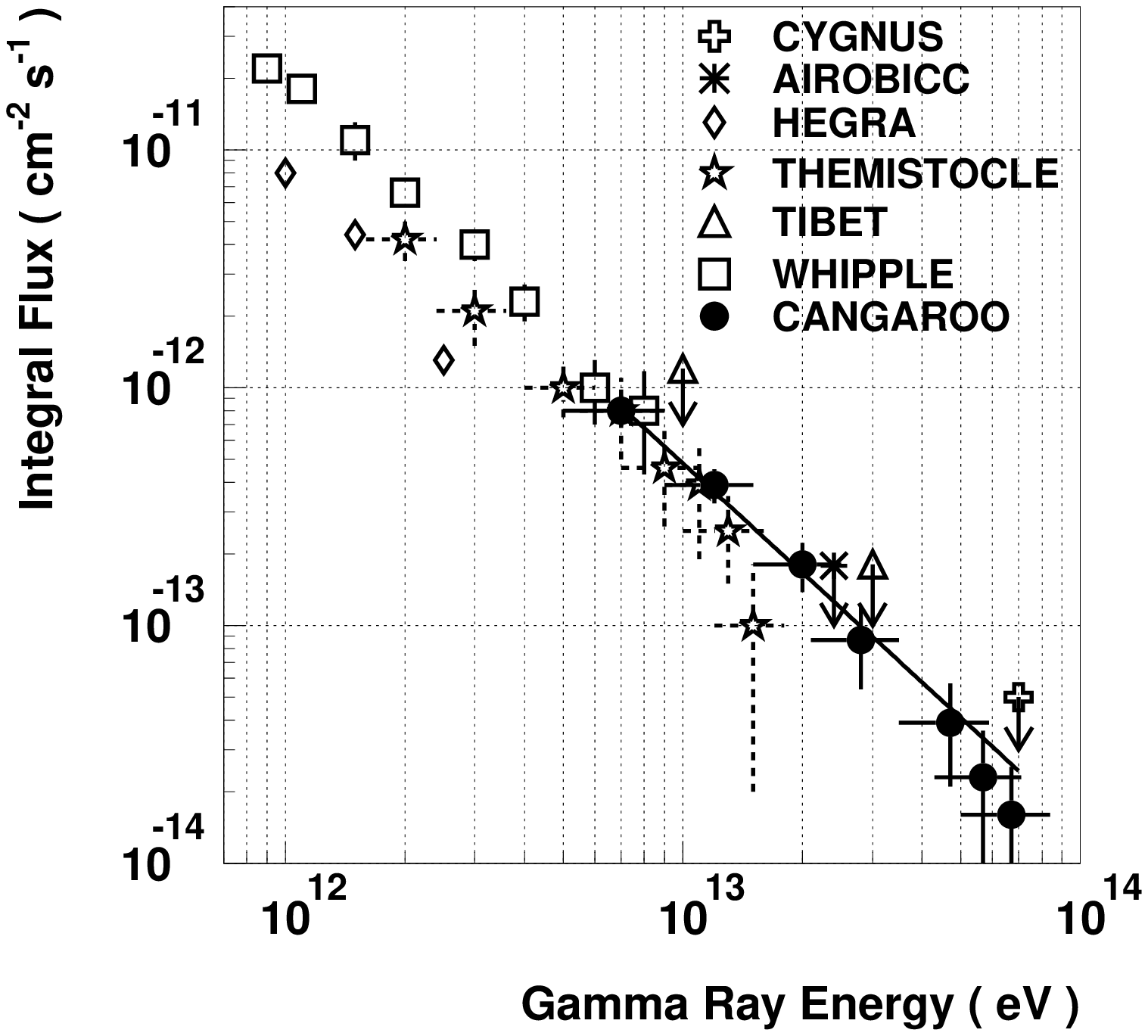]{
The integral spectra of the present result in comparison with
other experiments.
Recent results from the Whipple (Cater-Lewis et al.\ 1997) and 
the Themistocle (Djannati-Atai et al.\ 1995) groups 
are shown as well as upper limits from the 
Cygnus (Alexandreas et al.\ 1991), Tibet (Amenomori et al.\ 1992),
and AIROBICC (Karle et al.\ 1995).
The full line is the power law fit given by
$8.4\times 10^{-13}(E/{7\,\rm TeV})^{-1.53}
 \,\,{\rm cm}^{-2}{\rm s}^{-1}$.
\label{fig:f3}}

\figcaption[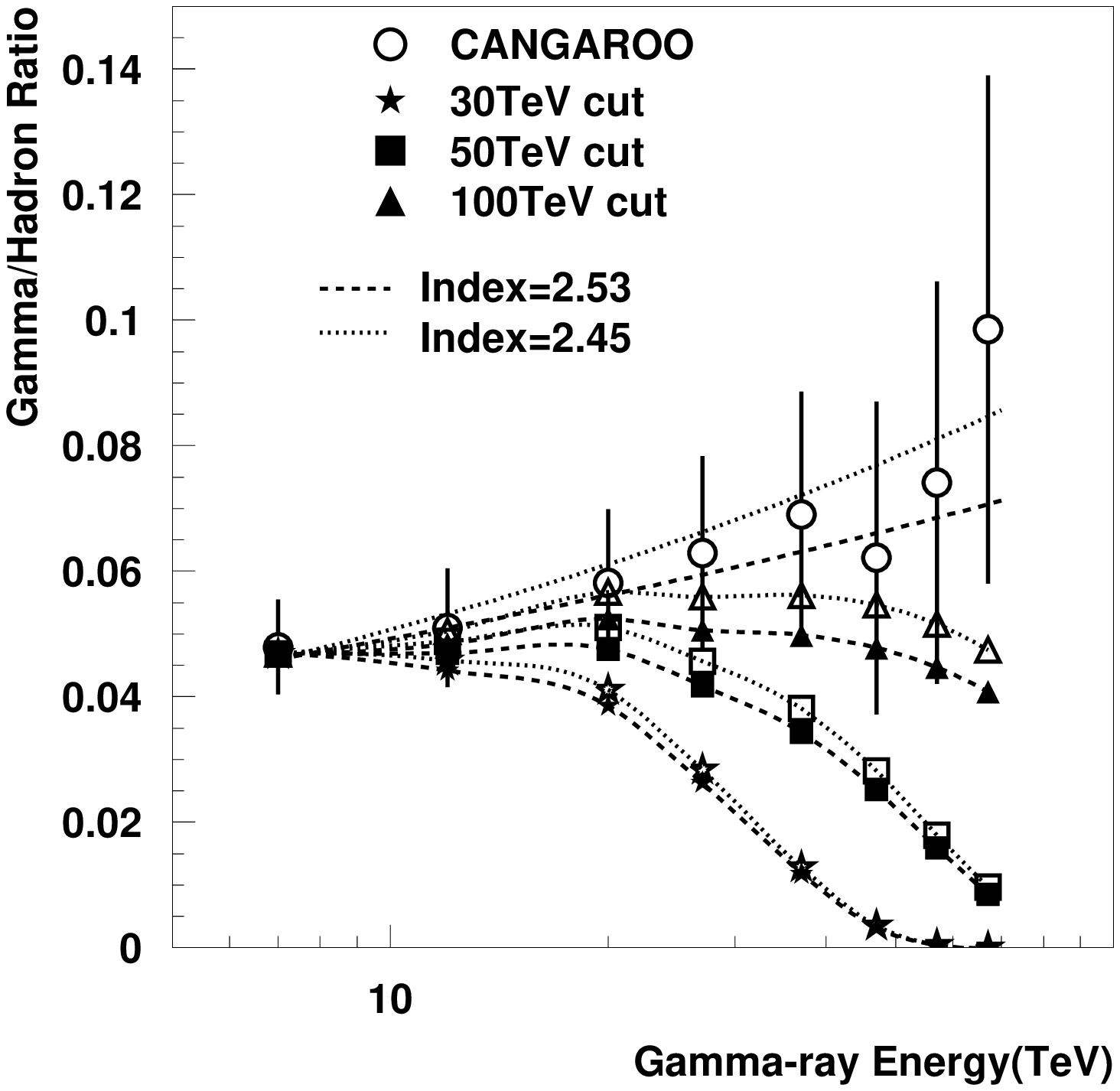]{
The ratio of the number of 
gamma rays to that of background events
as a function of threshold energy of gamma rays,
where gamma-rays events with $\alpha \le 15\arcdeg$ 
and background events with  (30\arcdeg--90\arcdeg) were used.
Simulated energy dependences of the ratio are also potted,
assuming the spectral cut-offs of 30(circle), 50(square)  and 100 TeV(triangle).
All simulated results are normarized to the observed ratio at 7 TeV.
Here  both differential indices
of ours (--2.53, broken line) and the Whipple group 
(--2.45, dotted line) were examined.
\label{fig:f4}}
\clearpage
\begin{figure}
\plotone{fig1.eps}
\end{figure}

\clearpage
\begin{figure}
\plotone{fig2.eps}
\end{figure}

\clearpage
\begin{figure}
\plotone{fig3.eps}
\end{figure}

\clearpage
\begin{figure}
\plotone{fig4.eps}
\end{figure}

\end{document}